\documentclass[]{aa}
\usepackage{natbib}
\usepackage{epsfig}
\def\ut#1{\mathop{\vtop{\ialign{##\crcr
     $\hfil\displaystyle{#1}\hfil$\crcr\noalign
     {\kern1pt\nointerlineskip}\hbox{$\hfil\sim\hfil$}\crcr
     \noalign{\kern1pt}}}}}

\def\undersymbol#1#2{\mathop{\vtop{\ialign{##\crcr
     $\hfil\displaystyle{#2}\hfil$\crcr\noalign
     {\kern1pt\nointerlineskip}\hbox{$\hfil#1\hfil$}\crcr
     \noalign{\kern1pt}}}}}

\begin{document}
\title{
Influence of magnification threshold on pixel lensing optical
depth, event rate and time scale distributions towards M31}
\author{F. De Paolis \inst{1},
        G. Ingrosso\inst{1},
        A.A. Nucita\inst{1},
        A.F. Zakharov\inst{2,3}
 }

\offprints{G. Ingrosso, \email{ingrosso@le.infn.it}} \institute{
Dipartimento di Fisica, Universit\`a di Lecce and INFN, Sezione di
Lecce, CP 193, I-73100, Italy
 \and
Institute of Theoretical and Experimental Physics,
           25, B.Cheremushkinskaya st., Moscow, 117259, Russia
 \and
 Astro Space
Centre of Lebedev Physics Institute, Moscow
\\
}
\authorrunning{De Paolis et al.}
\titlerunning{Magnification threshold influence on pixel lensing}
\date{}

\abstract{Pixel lensing is the gravitational microlensing of light
from unresolved stars contributing to the luminosity flux
collected by a single pixel. A star must be sufficiently
magnified, that is, the lens impact parameter must be less than a
threshold value $u_T$ if the excess photon flux in a pixel is to
be detected over the background. Assuming the parameters of the
Isaac Newton Telescope and typical observing conditions, we
present maps in the sky plane towards M31 of threshold impact
parameter, optical depth, event number and event time scale,
analyzing in particular how these quantities depend on $u_T$ in
pixel lensing searches. We use an analytical approach consisting
of  averaging on $u_T$ and the star column density the optical
depth, microlensing rate and event duration time scale. An overall
decrease in the expected optical depth and event number with
respect to the classical microlensing results is found,
particularly towards the high luminosity M31 inner regions. As
expected, pixel lensing events towards the inner region of M31 are
mostly due to self-lensing, while in the outer region dark events
dominate even for a 20\% MACHO halo fraction. We also find a
far-disk/near-disk asymmetry in the expected event number, smaller
than that found by \cite{Kerins04}. Both for self and dark lensing
events, the pixel lensing time scale we obtain is $\simeq 1 - 7$
days, dark events lasting roughly twice as long as self-lensing
events. The shortest events are found to occur towards the M31
South Semisphere. We also note that the pixel lensing results
depend on $\langle u_T \rangle$ and $\langle u_T^{~2} \rangle$
values and ultimately on the observing conditions and telescope
capabilities.

}

\maketitle

\section{Introduction}

Pixel lensing surveys towards M31 \citep{Crotts92,Baillon93} can
give valuable information to probe the nature of MACHOs (Massive
Astrophysical Compact Halo Objects) discovered in microlensing
experiments towards the LMC and SMC (Large and Small Magellanic
Clouds) \citep{Alcock93,Aubourg93} and also address the question
of the fraction of halo dark matter in the form of MACHOs in
spiral galaxies \citep{Alcock00}.

This may be possible due to both the increase in the number of
expected events and because the M31 disk is highly inclined with
respect to the line of sight and so microlensing by MACHOs
distributed in a roughly spherical M31 halo give rise to an
unambiguous signature: an excess of events on the far side of the
M31 disk relative to the near side \citep{Crotts92}.

Moreover, M31 surveys probe the MACHO distribution in a different
direction to the LMC and SMC and observations are made from the
North Earth hemisphere, probing the entire halo extension.

The Pixel lensing technique studies the gravitational microlensing
of unresolved stars \citep{Ansari97}. In a dense field of stars,
many of them contribute to each pixel. However, if one unresolved
star is sufficiently magnified, the increase of the total flux
will be large enough to be detected. Therefore, instead of
monitoring individual stars as in classical microlensing, one
follows the luminosity intensity of each pixel in the image. When
a significative (above the background and the pixel noise) photon
number excess repeatedly occurs, it is attributed to an ongoing
microlensing event if the pixel luminosity curve follows (as a
function of time) a Paczynski like curve \citep{Paczynski86}.

Clearly, variable stars could mimic a microlensing curve. These
events can be recognized by performing observations in several
spectral bands and monitoring the signal from the same pixel for
several observing seasons to identify the source.

Two collaborations, MEGA (preceded by the VATT/Columbia survey)
and AGAPE have produced a number of microlensing event candidates,
which show a rise in pixel luminosity in M31 (\citep{Crotts96,
Ansari99,Auriere01,Calchi02}).

More recently, based on observations with the Isaac Newton
Telescope on La Palma \citep{Kerins01}, the MEGA \citep{Mega04},
POINT-AGAPE \citep{Paulin03,Calchi03,Uglesich04} and WeCAPP
\citep{Wecapp03} collaborations claimed to find evidence of
several microlensing events.

In particular, the MEGA Collaboration \citep{Mega04} presented the
first 14 M31 candidate microlensing events, 12 of which are new
and 2 that have been reported by the POINT-AGAPE Collaboration
\citep{Paulin03}. The preliminary analysis of the spatial and
timescale distribution of the events supports their microlensing
nature. In particular the far-disk/near-disk asymmetry, although
not highly significant, is suggestive of the presence of an M31
dark halo.

The POINT-AGAPE Collaboration found in total a subset of four
short timescale, high signal-to-noise ratio microlensing
candidates, one of which is almost certainly due to a stellar lens
in the bulge of M31 and the other three candidates can be
explained either by stars in M31 and M32, or by MACHOs.

\begin{table*}
\caption{Parameters for the four M31 models considered by
\cite{Kerins04}. Columns are the model name, the component name,
the mass of the component, its central density $\rho_0$ and the
adopted cut-off radius $R$. Additional columns give, where
appropriate, the core radius $a$, the disk scale length $h$ and
height $H$, the flattening parameter $q$ and the B-band
mass-to-light ratio $M/L_B$ in solar units. }
\medskip
\begin{tabular}{c|ccccccccc}
\hline
Model & Component & Mass   & $\rho_0$ & $R$     & $a$     &  $h$    & $H$     & q & M/L$_B$                                \\
      && ($\times 10^{10}~M_{\odot}$)& ($\times M_\odot$ pc$^{-3}$) & (kpc) & (kpc) & (kpc) & (kpc) &   & \\
\hline
Massive halo & halo      & 191 &  0.25  & 155 & 2 & - & -  & 1    & -   \\
             & bulge     & 4.4 &  4.5   & 40  & 1 & - & -  & 0.6  & 9   \\
             & disk      & 3.2 & 0.24   & 40  & - & 6 &0.3 & -    & 4.5 \\
\hline
Massive bulge& halo      & 89 &  0.01   & 85 & 10 & - & -  & 1    & -   \\
             & bulge     & 8  &  2.5    & 40 & 1.5& - & -  & 0.6  & 14  \\
             & disk      & 11 &  0.6    & 40 & -  & 7 &0.3 & -    & 18  \\
\hline
Massive disk & halo      & 79 &  0.01   & 110 & 8 & - & -  & 1    & -  \\
             & bulge     & 4.4 &  4.5   & 40  & 1 & - & -  & 0.6  & 9  \\
             & disk      & 19 & 1.4     & 40  & - & 6 &0.3 & -    & 26 \\
\hline
Reference    & halo      & 123 &  0.065 & 100 & 4 & -   & -  & 1  & -  \\
             & bulge     & 4.4 &  4.5   & 40  & 1 & -   & -  & 0.6& 9  \\
             & disk      & 5.3 & 0.35   & 40  & - & 6.4 &0.3 & -  & 8  \\
\hline
\end{tabular}
\label{table0}
\end{table*}

In pixel lensing surveys, although all stars contributing to the same pixel
are candidates for a microlensing event, only the brightest stars
(usually blue and red giants) will be magnified enough to be
detectable above background fluctuations (unless for very high amplification
of main sequence stars, which are very unlikely events).

First evaluations have shown that the pixel lensing technique
towards M31 may give rise to a significant number of events due to
the large number of stars contributing to the same pixel
\citep{Baillon93,Jetzer94,Gould94,Colley95,Han96}.

Although these analytic estimates may be very rough, they provide
useful qualitative insights. To have reliable estimates in true
observational conditions one should use Monte-Carlo simulations
\citep{Ansari97,Kerins01}. In this way, given the capabilities of
the telescope and CCD camera used, the observing campaign and
weather conditions, one can estimate the event detection
efficiency as a function of event duration and maximum
amplification.

This study will be done in a forthcoming paper \citep{montecarlo}
with the aim of investigating the lens nature (i.e. the population
to which the lens belongs) for the events discovered by MEGA
\citep{Mega04} and POINT-AGAPE \citep{Paulin03}.

In this paper, instead of using Monte-Carlo simulations, we
estimate the relevant pixel lensing quantities by analyzing the
effect of the presence of a magnification threshold (or,
equivalently, of a threshold impact parameter $u_T$) in pixel
lensing searches towards M31. We use an analytic procedure
consisting of averaging the classical optical depth, microlensing
rate and event duration time scale on $u_T$, which depends on the
magnitude of the source being magnified.

The paper is organized as follows. In Section 2 we briefly discuss
the source - bulge and disk stars in M31 - and lens - stars in M31
and in the Milky Way (MW) disk, MACHOs in M31 and MW halos -
models we use. In Section 3 we discuss the pixel lensing
technique. In Sections 4 and 5 we present maps of optical depth,
event rate and typical event time duration, addressing the
modification with respect to classical microlensing values, due to
the influence of the threshold magnification in pixel lensing
searches. Finally in Section 6 we present some conclusions.

\section{Source and lens models}

The M31 disk, bulge and halo mass distributions are described
adopting the parameters of the Reference model in \cite{Kerins04},
which provides remarkably good fits to the M31 surface brightness
and rotation curve profiles.

This model, by using an average set of parameter values less
extreme with respect to the massive halo, massive bulge and
massive disk models in Table \ref{table0}, can be considered a
more likely candidate model for the mass distributions in the M31
galaxy.

Accordingly,
the mass density of the M31 disk stars is described by
a sech-squared profile
\begin{equation}
\rho_D (R,z) = \rho_D(0) \exp(-R/h)~ {\rm sech}^2(z/H),
\label{discoprofile}
\end{equation}
where  $R$ is the distance on the M31 disk plane, $H \simeq 0.3 $
kpc and $h\simeq 6.4 $ kpc are, respectively, the scale height and
scale lengths of the disk and $\rho_D(0)  \simeq 3.5 \times 10^8~
M_{\odot}$ kpc$^{-3}$ is the central mass density. The disk is
truncated at $R \simeq 40 $ kpc so that the total mass is $5.3
\times 10^{10}~M_{\odot}$.

As usual, the M31 disk is assumed to be inclined at the angle
$i=77^{0}$  and the azimuthal angle relative to the near minor
axis $\phi = -38.6^{0}$.

The M31 bulge is parameterized by a flattened power law of the
form
\begin{equation}
\rho_B(R,z) = \rho_B(0) \left[ 1+ \left( \frac {R}{a}\right)^2 +q^{-2}
\left( \frac {z}{a}\right)^2\right]^{-s/2}~,
\end{equation}
where the coordinates
$x$ and $y$ span the M31 disk plane ($z$ is perpendicular to it),
$\rho_B(0) \simeq 4.5 \times 10^9~M_{\odot}$ kpc$^{-3}$,
$ q \simeq 0.6$ is the ratio of the minor to major axes,
$a \simeq $ 1 kpc and $s \simeq 3.8$ \citep{Reitzel}.
The bulge is truncated at 40 kpc and its total mass
is $\simeq 4.4 \times 10^{10}~M_{\odot}$.

The dark matter in the M31 halo is assumed to follow an isothermal
profile
\begin{equation}
\rho_H(r) = \rho_H(0) \frac{a^2}{a^2+r^2}~,
\label{haloprofile}
\end{equation}
with core radius $a \simeq 4$ kpc, and central dark matter density
$\rho_H(0) \simeq 6.5 \times 10^{7}~M_{\odot}$ kpc$^{-3}$, so that
the total rotational velocity in the M31 halo is $v_{rot} \simeq
235$ km s$^{-1}$. The M31 halo is truncated at $R \simeq 100$ kpc
and the total dark mass within this distance is $\simeq 1.23
\times 10^{12}~M_{\odot}$.

As usual, the mass density profile for a MW disk is described with
a double exponential profile
\begin{equation}
\rho_D (R,z) = \rho_D(R_0) \exp(-(R-R_0)/h)~\exp(-|z|/H)~,
\end{equation}
with the Earth's position from the Galactic center at $R_0 \simeq
8.5$ kpc, scale height $H \simeq 0.3$ kpc, scale length $h \simeq
3.5$ kpc and local mass density $\rho_D(R_0) \simeq  1.67 \times
10^8~M_{\odot}$ kpc$^{-3}$.

The MW bulge
\footnote{
Although it does not contribute to microlensing towards M31, it
contributes to the dynamical mass of the Galaxy.}
is described by the triaxial
bulge model with mass density profile \citep{Dwek95}
\begin{equation}
\rho_B (x,y,z) = \frac{M_B}{ 8 \pi a b c } e^{-s^2/2}~,
\end{equation}
where $s^4 = [(x/a)^2+(y/b)^2]^2+(z/c)^4$, the bulge mass is $M_B
\simeq 2 \times 10^{10}~M_{\odot}$ and the scale lengths are $a
\simeq 1.49$ kpc, $b \simeq 0.58$ kpc and $c \simeq 0.40$ kpc.
Here, the coordinates $x$ and $y$ span the Galactic disk plane,
whereas $z$ is perpendicular to it.

The dark halo in our Galaxy is also assumed to follow an
isothermal profile
with core radius $a \simeq 5.6$ kpc and local dark matter density
$\rho_H(R_0) \simeq 1.09 \times 10^7~M_{\odot}$ kpc$^{-3}$.
The corresponding total asymptotic rotational velocity
is $v_{rot} \simeq 220$ km s$^{-1}$. The MW halo is truncated
at $R \simeq 100 $ kpc and the dark mass within this distance is
$\simeq 1.30 \times 10^{12}~M_{\odot}$.

For both M31 and MW halos, the fraction of dark matter
in form of MACHOs is assumed to be $f_{MACHO} \simeq 0.2$ \citep{Alcock00}.

Moreover, as usual, we assume the random velocities of stars and
MACHOs to follow Maxwellian distributions with one-dimensional
velocity dispersion $\sigma = 30, 100, 166$ km s$^{-1}$ and $30,
156$ km s$^{-1}$ for the M31 disk, bulge, halo and MW disk and
halo, respectively (see also \cite{Kerins01,An}). In addition an
M31 bulge rotational velocity of 30 km s$^{-1}$ is assumed.

\section{Pixel lensing basics }

\begin{figure}[htbp]
\vspace{9.0cm} \includegraphics{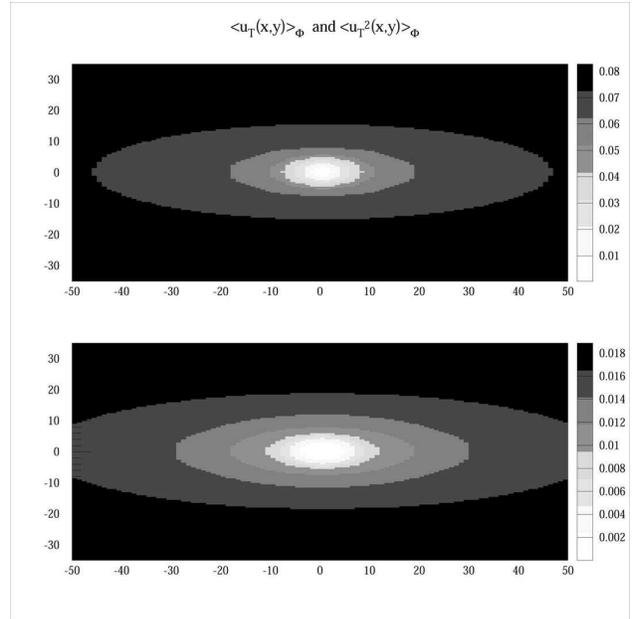} \caption{The mean impact parameter maps
$\langle u_T(x,y)\rangle _{\phi}$ and $ \langle
u_T^{~2}(x,y)\rangle _{\phi}$ towards M31 are given, for selected
directions towards M31 corresponding to different $(x,y)$
coordinates (in units of arcmin) centered on M31 and aligned along
the major and minor axes of the projected light profile. }
\label{fig1}
\end{figure}

\begin{figure}[htbp]
\vspace{7.5cm} \includegraphics{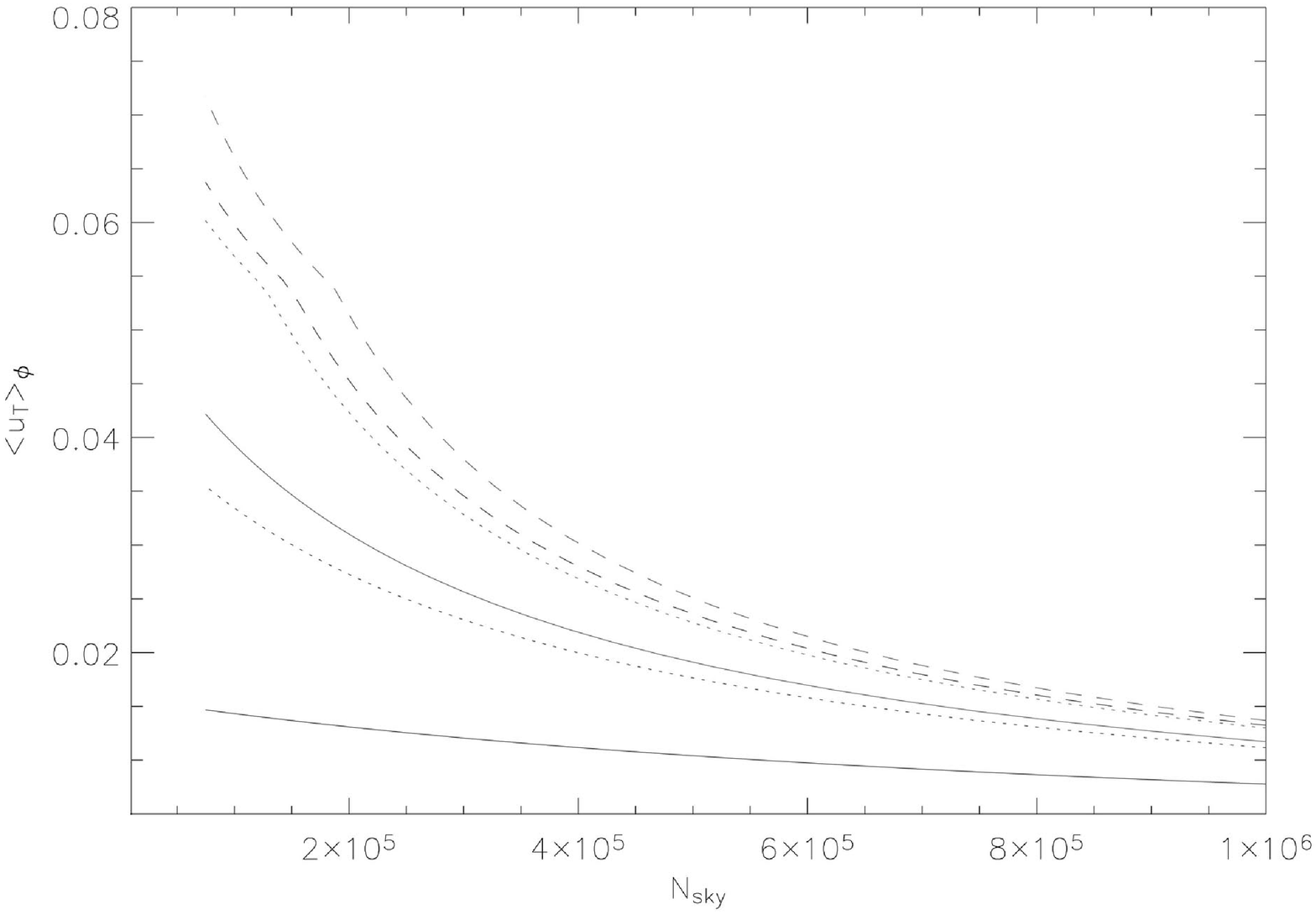} \caption{The mean impact parameter
$\langle u_T(x,y) \rangle _{\phi}$ is given as a function of the
background photon counts $N_{\rm sky}$ in a pixel, for selected
directions towards M31 corresponding to different $(x,y)$
coordinates (in units of arcmin) in the sky plane. Thin lines,
from the bottom to the upper part of the figure, refer to $(8,0)$,
$(16,0)$ and $(32,0)$; thick lines are for $(0,-2)$, $(0,-4)$ and
$(0,-8)$ coordinates, respectively. } \label{fig0}
\end{figure}

The main difference between gravitational microlensing and pixel
lensing observations relies in the fact that in pixel lensing a
large number of stars contribute to the same pixel and therefore
only bright and sufficiently magnified sources can be identified
as microlensing events. In pixel lensing analysis one usually
defines a minimum amplification that depends on the baseline
photon counts \citep{Ansari97}
\begin{equation}
N_{\rm bl} = N_{\rm gal}+N_{\rm sky}~,
\label{Nbl}
\end{equation}
which is the sum of the M31 surface brightness and sky contribution.

The excess photon counts per pixel due to an ongoing  microlensing event is
\begin{equation}
\Delta N_{\rm pix} = N_{\rm bl} [A_{\rm pix}-1] = f_{\rm see} N_s [A(t)-1]~,
\end{equation}
where $N_s$ and $N_{\rm bl}$ are the source and baseline photon
counts in the absence of lensing, $A(t)$ is the source
magnification factor due to lensing and $f_{\rm see}$ is the
fraction of the seeing disk contained in a pixel.

As usual, the amplification factor is given by (see, e. g.,
\cite{Griest} and references therein)
\begin{equation}
A(u) = \frac{u^2+2} {\sqrt{u^2(u^2+4)}}~,
\label{amplification}
\end{equation}
where
\begin{equation}
u(t)^2 = u_0^2 + \left( \frac{t-t_0}{t_E} \right)^2
\end{equation}
is the impact distance in units of the Einstein radius
\begin{equation}
R_E=[(4Gm_l/c^2) ~ D_l (D_s-D_l)/D_s]^{1/2}~,
\end{equation}
and $u_0$ is the impact parameter in units of $R_E$. Moreover,
$t_E=2R_E/v_{\perp}$ is the Einstein time, $t_0$ the time of
maximum magnification, $D_s$ and $D_l$ are the source and lens
distances from the observer and $v_{\perp}$ is the lens transverse
velocity with respect to the line of sight.

The number of photons in a pixel is given by
\begin{equation}
N_{\rm pix} =  \Delta N_{\rm pix} + N_{\rm bl}~,
\end{equation}
and a pixel lensing event will be detectable if the excess pixel
photon counts $\Delta N_{\rm pix}$ are greater than the pixel
noise
\begin{equation}
\sigma  = {\rm max} (\sigma_T, \sqrt{N_{\rm pix}})~,
\end{equation}
$\sigma_T$ being the minimum noise level determined by the pixel
flux stability and $\sqrt{N_{\rm pix}}$ the statistical photon
fluctuation.

By regarding a signal as being statistically significant
if it occurs at a level $3 \sigma$ above the baseline counts
$N_{\rm bl}$, one obtains
$A_{\rm min} \ge 1+3 \sigma/(f_{\rm see} N_s)$.
If $\sigma$ is taken equal to the minimum noise
level $\sigma_T$,
the obtained threshold magnification $A_T$ is \citep{Kerins01}
\begin{equation}
A_T = 1 + 0.0075 \frac{N_{\rm bl}}{f_{\rm see} N_s}
\end{equation}
which corresponds to a threshold value $u_T$ for the impact
distance via the relation in eq. (\ref{amplification}).
As one can see, $u_T$ depends on the source magnitude $M$,
the line of sight to M31 and the observing conditions.

\begin{figure*}[htbp]
\vspace{0.2cm}
\begin{center}
$\begin{array}{c@{\hspace{0.4in}}c} \epsfxsize=3.25in
\epsfysize=4.75in \epsffile{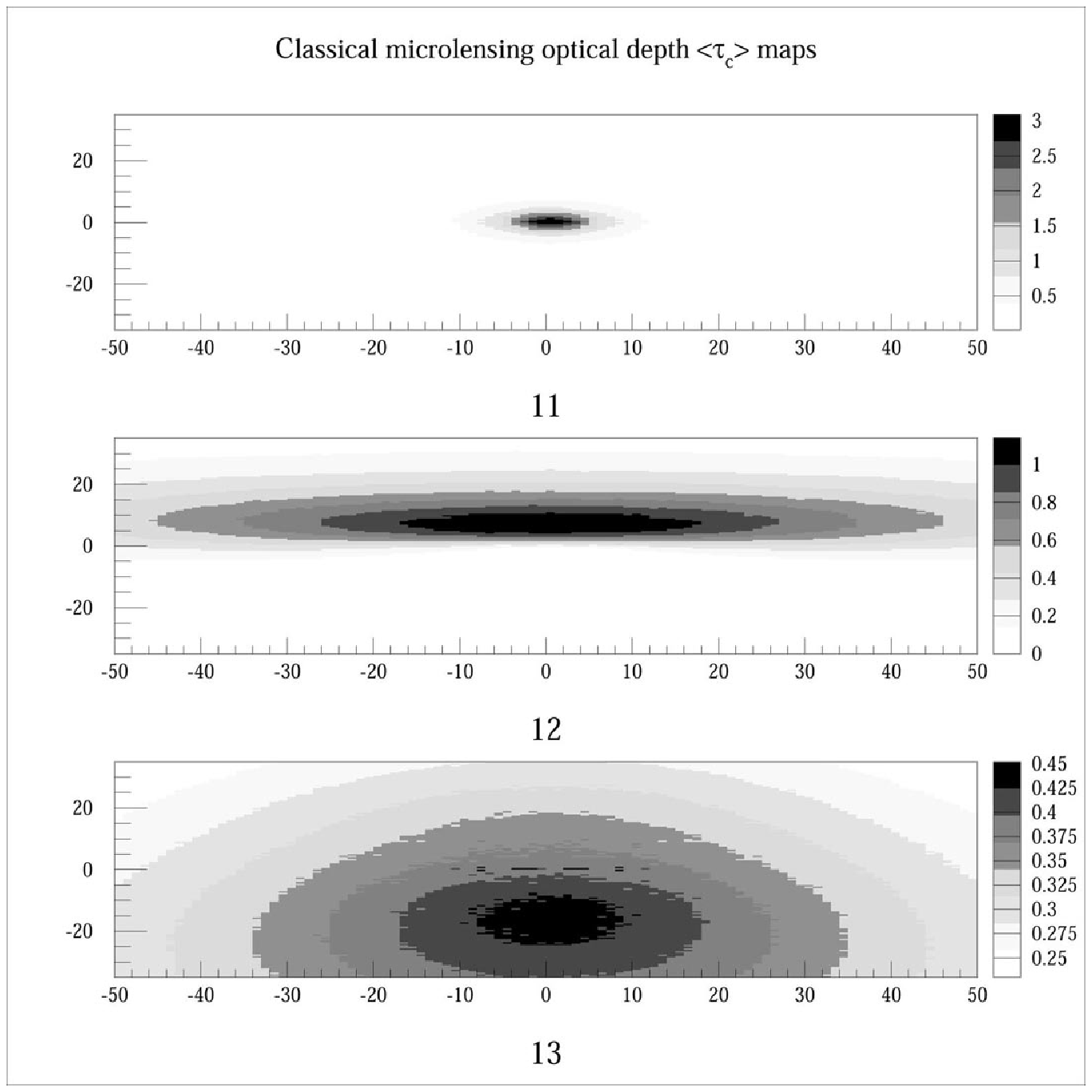} & \epsfxsize=3.25in
\epsfysize=4.75in \epsffile{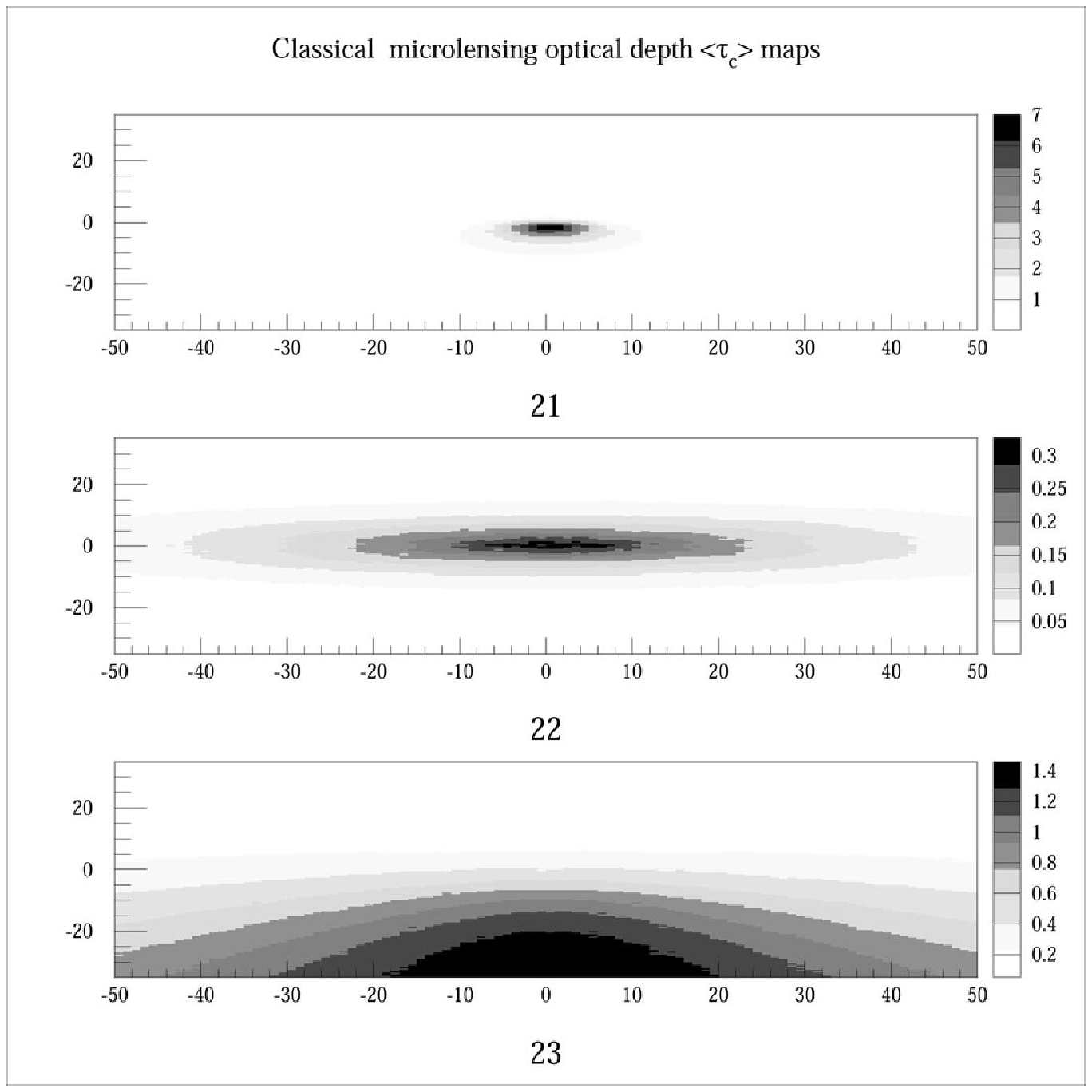}
\\ [0.cm]
\mbox{\bf a)} & \mbox{\bf b)}
\end{array}$
\end{center}
\caption{Mean optical depth $\langle \tau_c(x,y)\rangle $ maps (in
units of $10^{-6}$) towards M31 are given for selected source and
lens populations. The first index refers to the source stars (1
for M31 bulge, 2 for M31 disk) while the second one refers to the
lens populations (1 and 2 as above, 3 for MACHOs in the M31 halo).
Optical depth maps for lenses belonging to the MW disk and halo
populations are not given since the obtained results are almost
constant in any direction.}

\label{fig2}
\end{figure*}

\section{Pixel lensing optical depth and event rate}

In pixel lensing analysis, the effect of the existence of
the threshold magnification (or, equivalently, of a threshold value
of the impact parameter) is usually taken
into account in estimating the pixel lensing rate \citep{Kerins01,Kerins03}
\begin{equation}
\Gamma_p (x,y) = \langle u_T(x,y) \rangle _{\phi} ~ \Gamma_c (x,y)~,
\label{gammap}
\end{equation}
where $x$ and $y$ are coordinates in the plane orthogonal to the
line of sight and $\langle u_T(x,y) \rangle _{\phi}$ is averaged
on the source magnitude $M$, namely
\begin{equation}
\langle u_T(x,y) \rangle _{\phi} = \frac{ \int_{M_1}^{M_2} u_T(M;x,y) \phi(M) dM}
                         { \int_{M_1}^{M_2}            \phi(M) dM}~,
\end{equation}
$M_1-M_2$ (to be specified below) being the limiting values for
the source magnitude and $\phi(M) dM$ the luminosity function,
i.e. the number density of sources in the absolute magnitude
interval $(M,M+dM)$.

Pixel lensing event detection is usually performed in R or I bands
in order to minimize light absorption by the intervening dust
in M31 and MW disks. Indeed, these bands offer the best compromise between
sampling and sky background, while other bands (B and V) are commonly used
to test achromaticity of the candidate events.

In the present analysis, as reference values, we adopt the
parameters of the Sloan-r filter on the Wide-Field Camera of the
Isaac Newton Telescope \citep{Kerins01}. Therefore, since the red
giants are the most luminous stars in the red band,
we may safely assume that the overwhelming majority of the pixel
lensing event sources are red giants. Moreover, gives the lack of
precise information about the stellar luminosity function in the
M31 galaxy, we assume that the same function $\phi(M)$ holds both
for the Galaxy and M31 and does not depend on position.

Accordingly, in the range of magnitude $-6 \le M \le 16$ the
stellar luminosity function is proportional to the following
expression \citep{MamonSoneira}
\begin{equation}
\phi(M) \propto
\frac{ 10^{\beta(M-M^*)}  }
               { [ 1+10^{-(\alpha-\beta)\delta(M-M^*)}]^{1/\delta} }~,
\end{equation}
where, in the red band,
$M^*= 1.4$, $\alpha \simeq 0.74$, $\beta = 0.045$ and $\delta= 1/3$.

On the other hand, the fraction of red giants (over the total star number)
as a function of $M$
can be approximated as \citep{MamonSoneira}
\begin{eqnarray}
f_{RG}(M)
&=& 1 - C \exp[\alpha(M+\beta)^{\gamma}]~~~{\rm for}~~-6 \le M \le 3
\nonumber\\
     &=&  0~~~~~~~~~~~~~~~~~~~~~~~~~~~~~~{\rm for}~~M \ge 3~,
\end{eqnarray}
where, in the red band,
$C \simeq 0.31$, $\alpha \simeq 6.5 \times 10^{-4}$, $\beta = 7.5$
and $\gamma \simeq 3.2$. Therefore, the fraction of red giants
averaged over the magnitude is given by
\begin{equation}
 \langle f_{RG} \rangle  =
\frac { \int_{-6}^{3} \phi(M) f_{RG}(M) dM }
{\int_{-6}^{16} \phi(M) dM }~,
\end{equation}
from which we obtain
$\langle f_{RG} \rangle \simeq 5.3 \times 10^{-3}$.

\begin{figure*}[htbp]
\vspace{0.2cm}
\begin{center}
$\begin{array}{c@{\hspace{0.4in}}c} \epsfxsize=3.25in
\epsfysize=4.75in \epsffile{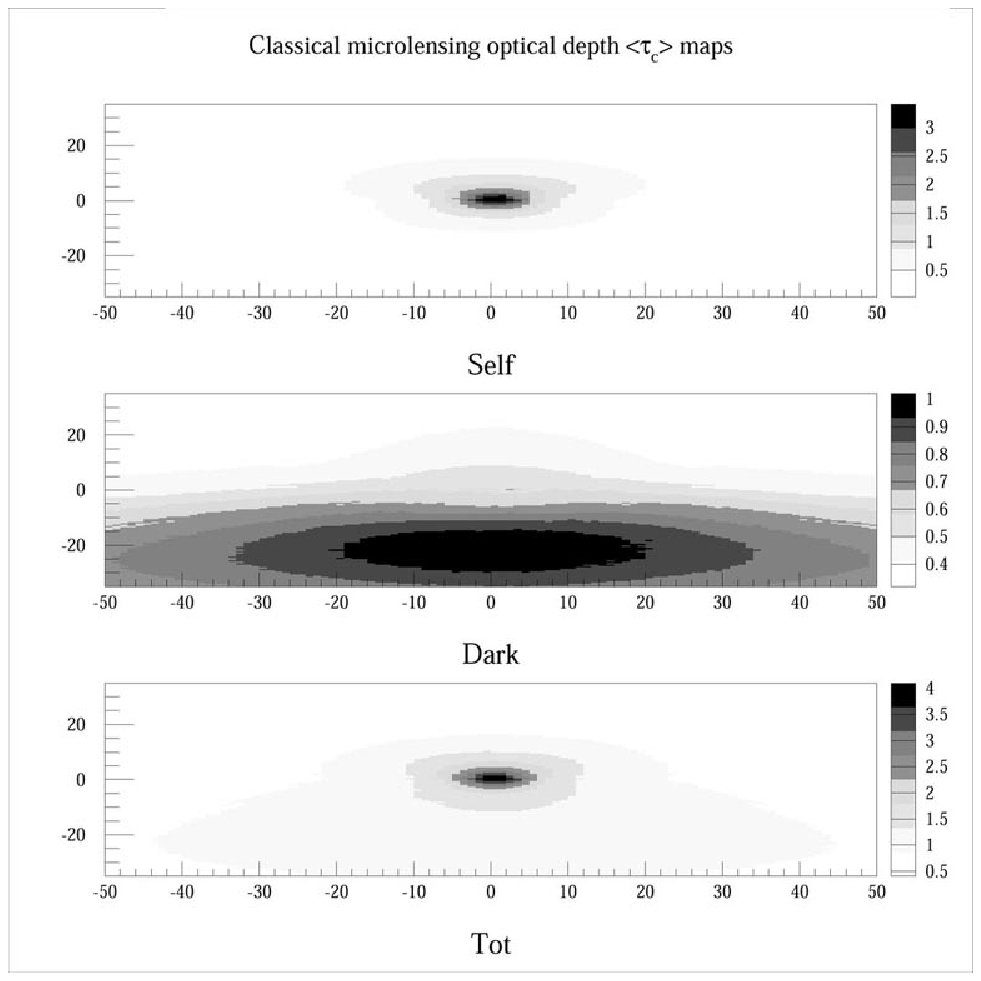} & \epsfxsize=3.25in
\epsfysize=4.75in \epsffile{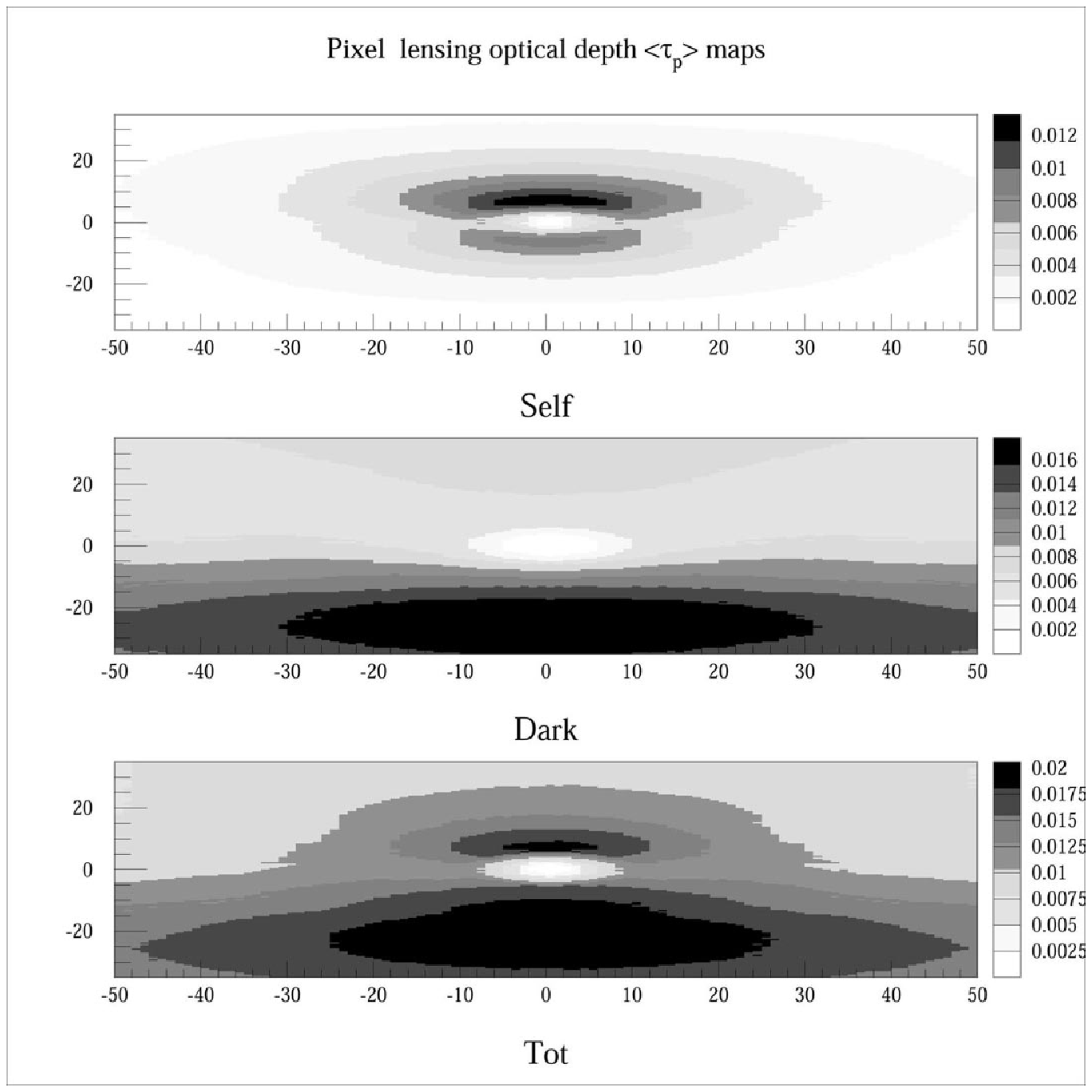}
\\ [0.cm]
\mbox{\bf a)} & \mbox{\bf b)}
\end{array}$
\end{center}
\caption{
In panel a), the mean classical optical depth
$\langle \tau_c(x,y)\rangle $ maps
(in units of $10^{-6}$) towards M31 are given for self, dark and total
lensing. In panel b), the mean pixel lensing optical depth
$\langle \tau_p(x,y)\rangle$ maps
are given, in the same cases.
}
\label{fig3}
\end{figure*}

Averaging the pixel lensing rate in eq. (\ref{gammap})
on the source density we obtain
\begin{equation}
\langle \Gamma_p (x,y) \rangle  =
\langle u_T(x,y) \rangle _{\phi}~\langle \Gamma_c (x,y)\rangle _{n_s}~,
\label{eqno16}
\end{equation}
where the mean classical rate $\langle \Gamma_c (x,y) \rangle _{n_s}$
is
\begin{equation}
\langle \Gamma_c(x,y) \rangle _{n_s}
           = \frac{ \int \Gamma_c(D_s;x,y)~n_s(D_s;x,y) dD_s}
                  { \int                   n_s(D_s;x,y) dD_s}~.
\label{rateposition}
\end{equation}

\begin{figure*}[htbp]
\vspace{0.2cm}
\begin{center}
$\begin{array}{c@{\hspace{0.4in}}c} \epsfxsize=3.25in
\epsfysize=4.75in \epsffile{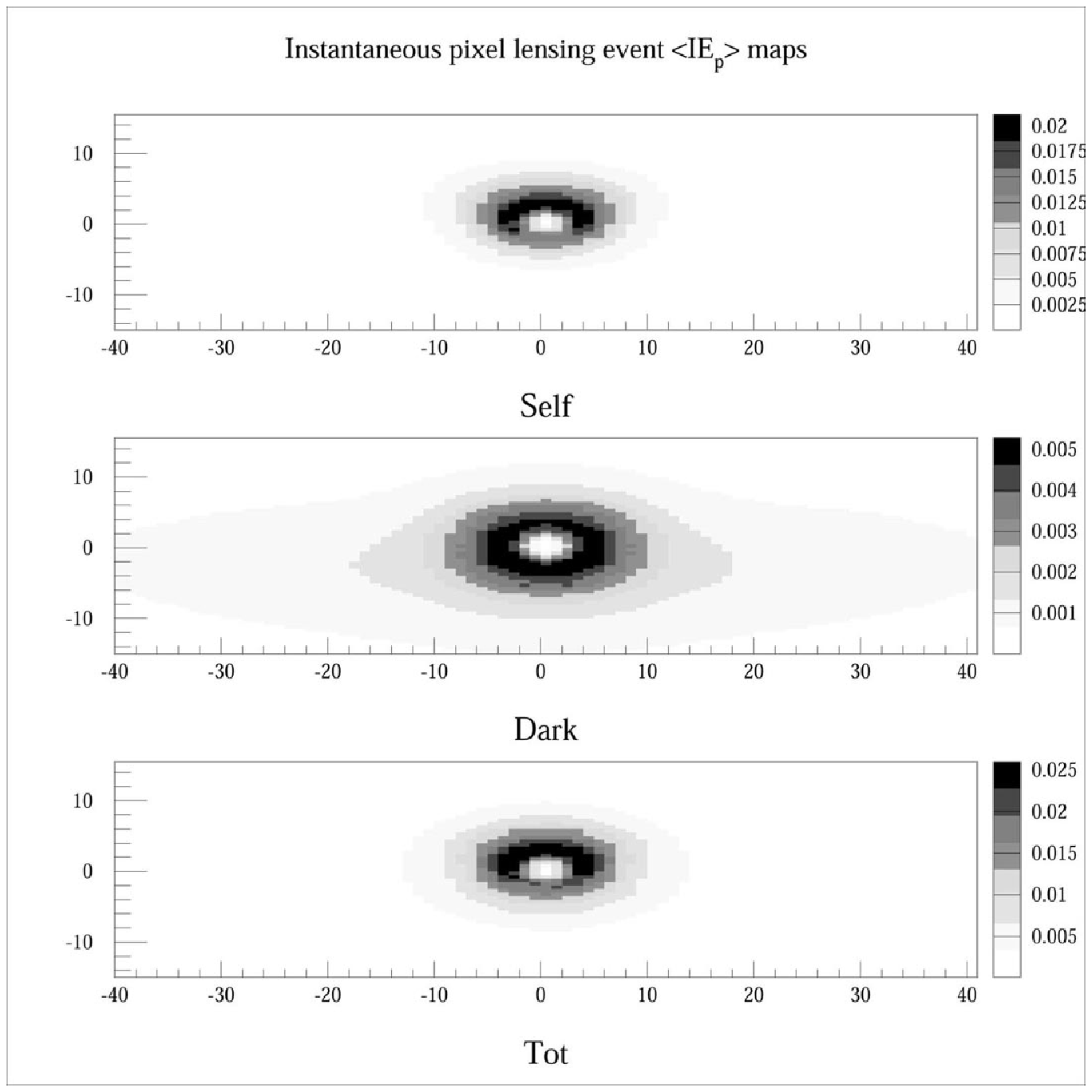} & \epsfxsize=3.25in
\epsfysize=4.75in \epsffile{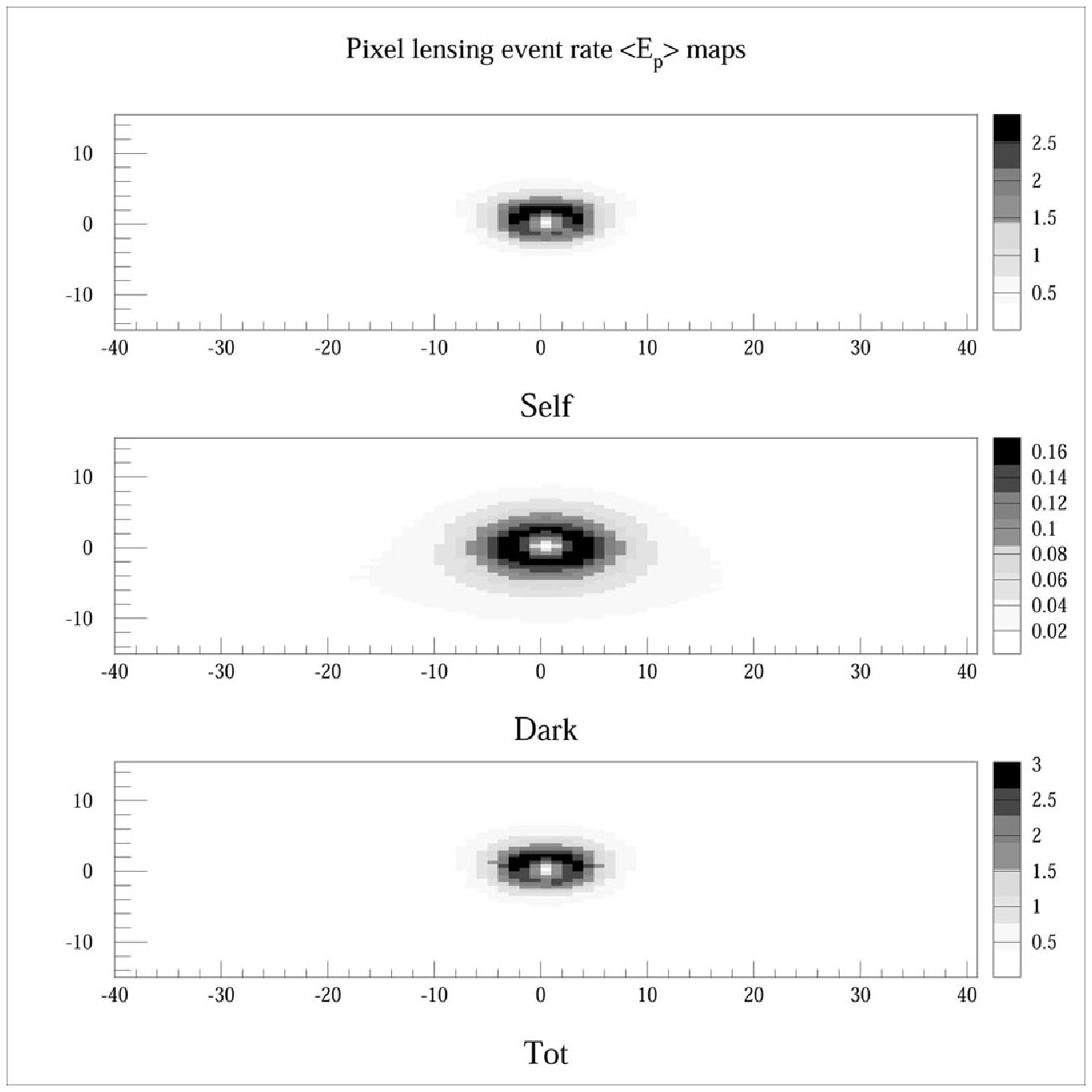}
\\ [0.cm]
\mbox{\bf a)} & \mbox{\bf b)}
\end{array}$
\end{center}
\caption{In panel a), the instantaneous pixel lensing event number density
$\langle IE_p(x,y)\rangle $ maps
(events per arcmin$^{2}$) towards M31 are given for self, dark and total
lensing.
In panel b) maps of pixel lensing event rate
$\langle E_p(x,y)\rangle$ (events per year and per arcmin$^{2}$) are given,
in the same cases.}
\label{fig5}
\end{figure*}

In turn, for a fixed source distance, $\Gamma_c(D_s;x,y)$ is
obtained by the classical microlensing rate for
a lens of mass $m_l$ \citep{Griest},
by averaging on the lens mass, namely
\begin{equation}
\Gamma_c(D_s;x,y)= \frac{\int \Gamma_c (m_l;D_s,x,y)~P(m_l) dm_l}
                           {\int P(m_l) dm_l }~,
\end{equation}
where $P(m_l)$ is the lens mass distribution function.

For lenses belonging to the bulge and disk star populations,
lenses are assumed to follow a broken power law
(see e.g. \cite{An} and references therein)
\begin{eqnarray}
P(m_l) &\propto&
m_l^{-1.4}~~~{\rm for}~~0.1~M_{\odot} \le m_l \le 0.5~M_{\odot}
\nonumber\\
&\propto&        m_l^{-2.2}~~~{\rm for}~~0.5~M_{\odot} \le m_l \le (m_l)_{up}
\end{eqnarray}
where the upper limit $(m_l)_{up}$ is $1~M_{\odot}$
for M31 bulge stars and $10~M_{\odot}$ for M31 and MW disk stars.
The resulting mean mass for lenses in the bulges and disks are
$\langle m_b \rangle \sim 0.31~M_{\odot}$ and
$\langle m_d \rangle \sim 0.53~M_{\odot}$, respectively.

For the lens mass in the M31 and MW halos we
assume the $\delta$-function approximation and we take
a MACHO mass
$m_{MACHO} \simeq 0.5~M_{\odot}$, according to the mean value
in the analysis of
microlensing data
towards LMC \citep{Alcock00}.

As usual, the mean number of expected events in classical microlensing
$\langle E_c(x,y) \rangle $
and pixel lensing $\langle E_p(x,y)\rangle $, respectively,
are related to the observation time $t_{obs}$,
source column density $N_s$ and mean fraction of red giants
$\langle f_{RG} \rangle $ by
\begin{equation}
\langle E_c(x,y) \rangle  = \langle \Gamma_c(x,y)\rangle _{n_s}~
t_{obs}~ N_{s}(x,y)~ \langle f_{RG} \rangle~,
\end{equation}
\begin{equation}
\langle E_p(x,y)\rangle  = \langle \Gamma_p(x,y)\rangle _{n_s}~
t_{obs}~ N_{s}(x,y)~ \langle f_{RG} \rangle~,
\end{equation}
where the source column density is
\begin{equation}
N_{s}(x,y) = \int n_s(D_s;x,y) dD_s~.
\end{equation}

However, the classical microlensing rate depends on several source
and lens parameters, in particular on the lens mass and transverse
velocity of the source and lens. Therefore, due to the parameter
degeneracy, it does not give precise information on the lens
population, at least in the M31 regions where microlensing by
stars in M31 itself (self-lensing) and by MACHOs in M31 and MW
halos (dark lensing) occur with comparable probability. Indeed, as
usual, the probability for a given lens population is
\begin{equation}
P_{l} = \frac{ \sum_s \Gamma_{sl}} {\sum _s \sum _l \Gamma_{sl}}~.
\label{pp}
\end{equation}

On the other hand, the classical microlensing optical depth
\begin{equation}
\tau_c (D_s;x,y)  = \int_0^{D_s} \pi R_E^2 n_l(D_l)~dD_l
\label{tauc}
\end{equation}
is a geometrical quantity and depends on a small number of
parameters and can be used as in eq. (\ref{pp}) to determine the
lens nature.

Physically the optical depth is the number of ongoing microlensing
events per source star at any instant in time. So, one can also
compute the instantaneous event number density, as a function of
position, by multiplying the optical depth  by the number density
of sources
\begin{equation}
\langle IE_c(x,y) \rangle =
\langle \tau_c(x,y)\rangle_{ns} ~ N_{s}(x,y)~ \langle f_{RG} \rangle~.
\label{evistc}
\end{equation}

However, eq. (\ref{tauc}) is the usual definition for the optical
depth in classical gravitational microlensing, while in the case
of pixel lensing it is necessary to take into account the effect
of $u_T(M;x,y)$.

In order to generalize the $\tau$ definition to the pixel lensing
case, a new definition (which joins the advantage of using a
geometrical quantity with the main characteristic of the pixel
lensing technique, i.e. the effect of the baseline) is introduced
(see also \cite{Kerins04})
\begin{equation}
  \langle \tau_p(x,y)\rangle  = \langle u_T^{~2}(x,y) \rangle _{\phi}
~ \langle \tau_c (x,y) \rangle _{n_s}~,
\label{taup}
\end{equation}
where $\langle \tau_c (x,y) \rangle _{n_s}$ is
\begin{equation}
\langle \tau_c(x,y)\rangle _{n_s}
              = \frac {\int \tau_c(x,y) ~ n_s(D_s;x,y) dD_s}
                      {\int             n_s(D_s;x,y) dD_s}~.
\label{tauposition}
\end{equation}
The factor $\langle u_T^{~2}(x,y) \rangle _{\phi}$ in eq.
(\ref{taup}) comes from the consideration that the Einstein radius
$R_E$, which enters quadratically in $\tau_c$, has to be
multiplied by $u_T$ (always less than unity for pixel lensing).

Accordingly, the instantaneous event number density in pixel lensing is
given by
\begin{equation}
\langle IE_p(x,y) \rangle =
\langle \tau_p(x,y)\rangle_{ns} ~ N_{s}(x,y)~ \langle f_{RG} \rangle~.
\label{evistp}
\end{equation}

Here we note that in evaluating $\langle E_p(x,y) \rangle$ for
each model considered in Table \ref{table0}, we have to take into
account that the number of detectable pixel lensing events does
not depends on {\bf the typical source luminosity $L_s$} to first
order \citep{Kerins04}. Indeed, although for a fixed source
luminosity $L_s$ the number of sources $N_s \propto L_s^{-1}$, the
pixel lensing rate per source $\Gamma_p \propto L_s$, \footnote{
Indeed, the faintest detectable pixel lensing events require a
threshold source magnification $A_T \propto L_s^{-1}$, to be seen
against the local background. On the other hand, in pixel lensing
the bulk of observed events involves highly magnified sources for
which $A_T \propto u_T^{~-1}$. Therefore since $\Gamma_p \propto
u_T$, it follows that $\Gamma_p \propto L_s$.} so that the event
number does not depend on $L_s$. The same also holds for the
instantaneous event number density $\langle IE_p(x,y) \rangle$.
\begin{figure*}[htbp]
\vspace{0.2cm}
\begin{center}
$\begin{array}{c@{\hspace{0.4in}}c} \epsfxsize=3.25in
\epsfysize=4.75in \epsffile{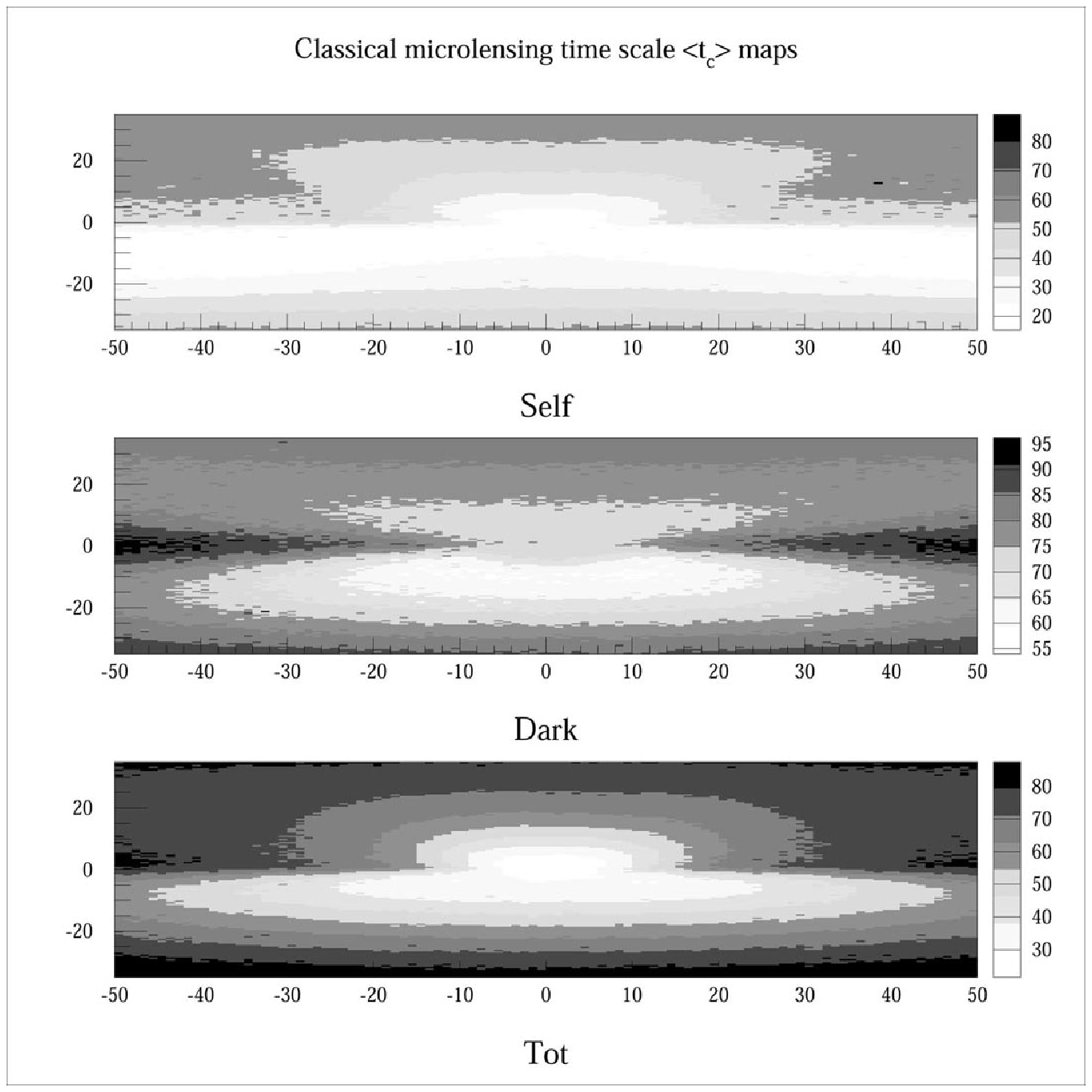} & \epsfxsize=3.25in
\epsfysize=4.75in \epsffile{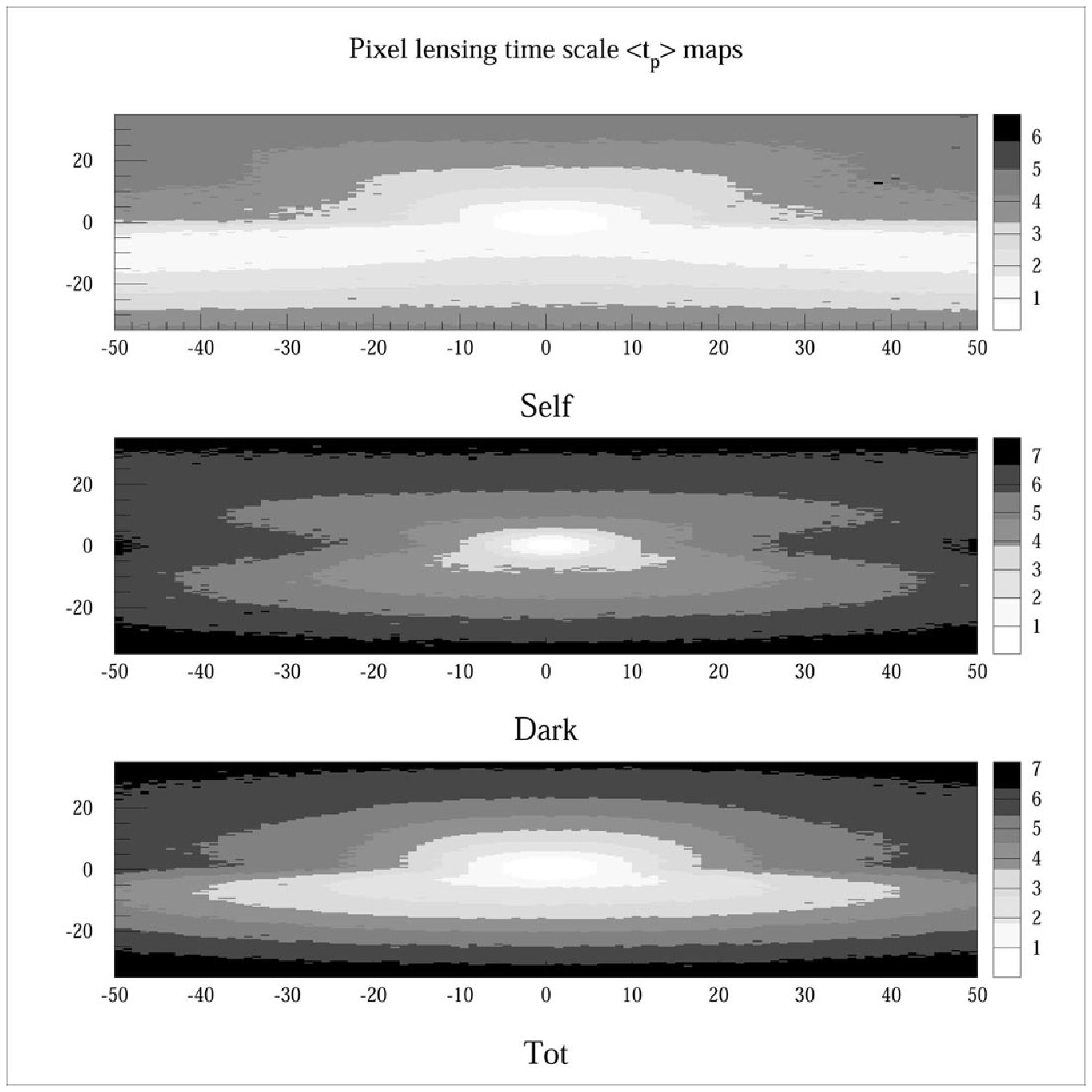}
\\ [0.cm]
\mbox{\bf a)} & \mbox{\bf b)}
\end{array}$
\end{center}
\caption{In panel a), mean classical event duration time $\langle
t_c(x,y)\rangle $ (in days) maps  towards M31 are given for self,
dark and total lensing. In panel b) for pixel lensing, maps of
$\langle t_p(x,y)\rangle $ are given, in the same cases.}
\label{fig6}
\end{figure*}

\section{Pixel lensing event duration}

In pixel microlensing, due to the large number of stars
simultaneously contributing to the same pixel, the flux from a
single star in the absence of lensing is generally not observable.
Thus, the Einstein time $t_E$ cannot be determined reliably by
fitting the observed light curve.

Indeed, another estimator of the event time duration has been
proposed, namely the full-width half-maximum event duration
$t_{FWHM}$, which depends on $t_E$ and $u_0$ \citep{Gondolo}
\begin{equation}
t_{FWHM} = t_E w(u_0)~,
\end{equation}
where $w(u_0)$ is given by
\begin{equation}
w(u_0) = 2 \sqrt{2 f [f(u_0^2)]-u_0^2}
\end{equation}
and $f(x) = A(x) - 1$,
where $A(x)$ is the amplification factor in eq. (\ref{amplification}).

This quantity can be put in a different form \citep{Kerins01}
\begin{equation}
t_{FWHM} = 2 \sqrt{2} t_E \left( \frac{a+2}{\sqrt{a^2+4a}} -
\frac{a+1}{\sqrt{a^2+2a}} \right)^{1/2}~,
\label{t12kerins01}
\end{equation}
where a = $A_{\rm max}-1$.

In the limit of large amplification $A_{\rm max} >> 1$ (or, equivalently,
$u_0 << 1$) one obtains
\begin{equation}
a \simeq \frac{3 u_0}{8} + \frac{1}{u_0} -1 + O(u_0^3)~.
\label{30}
\end{equation}
Using eq. (\ref{30}) in eq. (\ref{t12kerins01}),
the full-width half-maximum event duration can be
approximated by
\begin{equation}
t_{FWHM} \simeq 2 \sqrt{2}~ t_E~ u_0 ( 1-  u_0) + O(u_0^3) ~.
\label{t12appr}
\end{equation}

Clearly, while in classical microlensing $u_0$ may be determined,
in pixel microlensing the background overcomes the source baseline
making $u_0$ unknown, implying that an average procedure on $u_0$
is needed to estimate the mean event duration. Since the impact
parameter $u_0$ varies in the range $(0,u_T) $ and the probability
of $u_0$ being in the range $u_0$ -- $u_0+du_0$ is $P(u_0) du_0
\propto 2 \pi u_0 du_0$ (the area of the circular ring of radius
$u_0$ and thickness $du_0$), by averaging $t_{FWHM}$ on the impact
parameter, in the limit of large amplification, one gets
\begin{equation}
\langle t_{FWHM} \rangle _{u_0} \simeq 2 \sqrt{2} ~ t_E ~
   \frac {\int_0^{u_T} u_0^2(1-u_0)du_0}
         {\int_0^{u_T} u_0 du_0 }~ \simeq
          u_T~ t_c~,
\label{37}
\end{equation}
where
\begin{equation}
t_c = \frac{ 4 \sqrt{2}} {3} ~ t_E(x,y)~.
\end{equation}
Inspection of the relation in eq. (\ref{37}) and
of eqs. (\ref{gammap}), (\ref{eqno16})
and  (\ref{taup}),
lead us to introduce for pixel lensing a new event time scale estimator
$ \langle t_p(x,y) \rangle $ defined as
\begin{equation}
\langle t_p(x,y)\rangle  = \langle u_T(x,y) \rangle_{\phi}~
\langle t_c (x,y) \rangle_{ns}~,
\label{tempop}
\end{equation}
where
\begin{equation}
\langle t_c(x,y)\rangle _{n_s}
              = \frac {\int t_c(x,y) ~ n_s(D_s;x,y) dD_s}
                      {\int            n_s(D_s;x,y) dD_s}~,
\end{equation}
and
\begin{equation}
t_c(x,y) = \frac{ 4 \sqrt{2}} {3}
\frac{ \int t_E(x,y) d\Gamma(x,y)}{\Gamma(x,y)}~.
\end{equation}

Clearly, the pixel lensing time scale $\langle t_p(x,y)\rangle$
turns out to be the full-width half-maximum event duration
averaged over the impact parameter.

\section{Results}
Before discussing the results obtained, we summarize some
assumptions used in the present analysis. First, we assume perfect
sensitivity to pixel lensing event detection in M31 pixel lensing
searches. Moreover, as reference values, we use the parameters for
the Isaac Newton Telescope and WFC (Wide-Field Camera) adopted by
the POINT-AGAPE collaboration \citep{Kerins01,Kerins03}.

The Telescope diameter, the pixel field of view and the image
exposition time are $2.5$ m, $0.33$ arcsec and $t_{exp}= 760$ s,
respectively. We also assume a gain or conversion factor of 2.8
e$^{-}$/ADU, and a loss factor $\simeq 3$, both for atmospheric
and instrumental effects. The zero-point with the Sloan-r WFC is
$\sim 24.3$ mag arcsec$^{-2}$.
\begin{figure}[htbp]
\vspace{7.5cm} \includegraphics{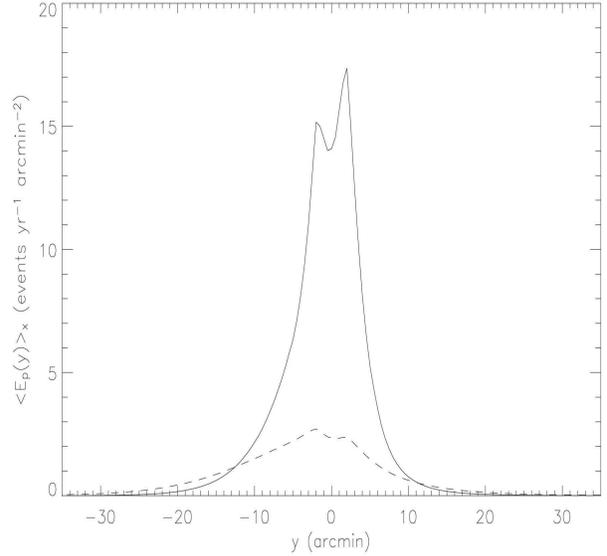} \caption{The projected (along the $x$
axis) mean event number $\langle E_p(y) \rangle _x$ is given as a
function of the coordinate $y$ for the Reference model. The dashed
line refers to dark lensing events by MACHOs in M31 and MW halos
while the solid line is for self-lensing events by stars in M31
bulge and disk. } \label{f8f8}
\end{figure}

To take into account the effect of seeing, we employ an analysis
based on superpixel photometry. Adopting a value of 2.4 arcsec for
the worst seeing value, we take a superpixel dimension of 7x7
pixel and adopt a minimum noise level of $\sigma_T \sim 2.5 \times
10^{-3} N_{\rm bl}$. We also assume that typically 87 per cent of
a point spread function (PSF) positioned at the center of a
superpixel is contained within the superpixel itself.
\begin{figure}[htbp]
\vspace{7.5cm} \includegraphics{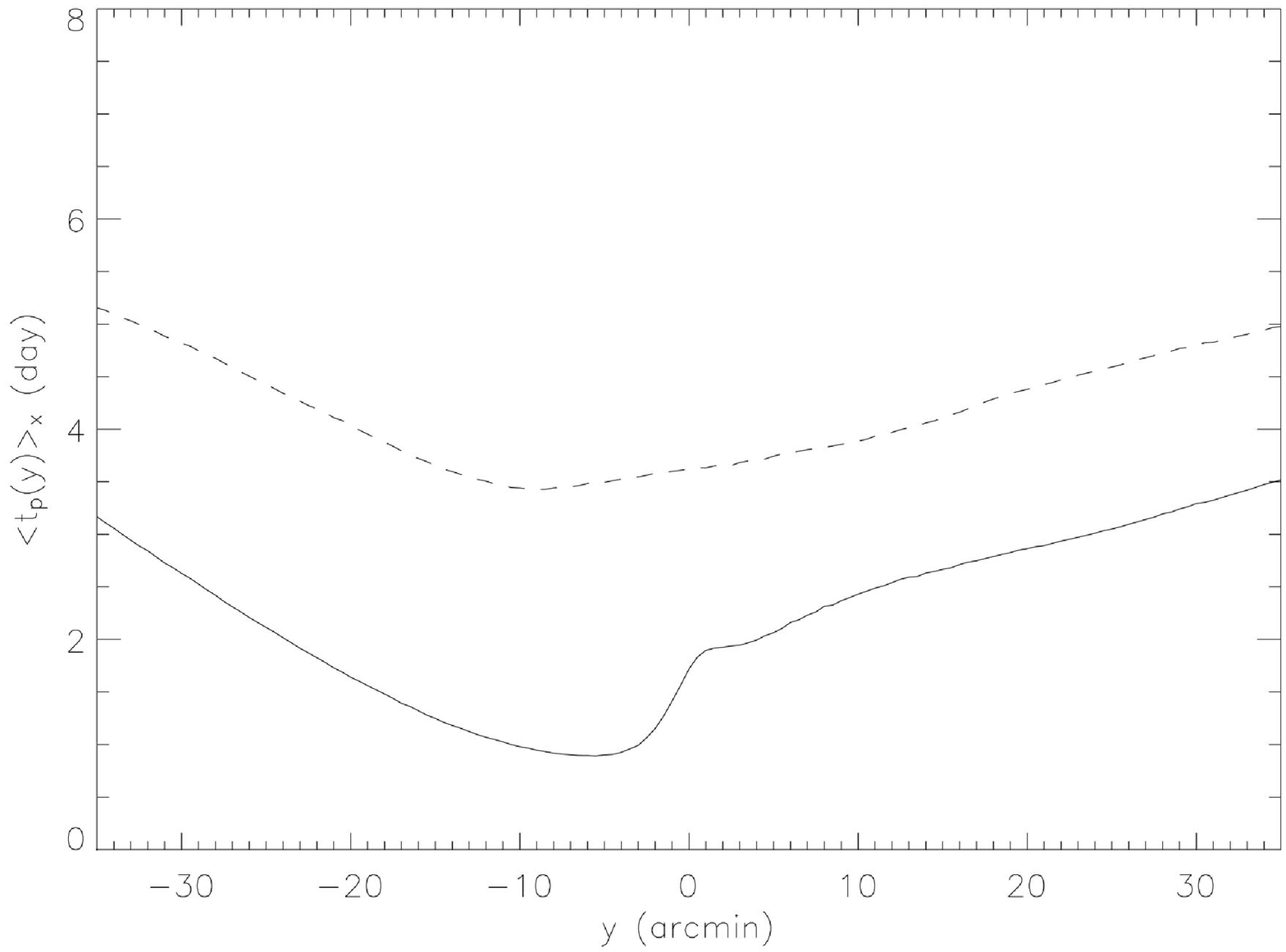} \caption{The pixel lensing event
duration $\langle t_p(y) \rangle _x$ averaged along the $x$
direction is given as a function of the $y$ coordinate. The dashed
line refers to dark lensing events by MACHOs in M31 and MW halos
while the solid line is for self-lensing events by stars in M31
bulge and disk. } \label{f9f9}
\end{figure}

The considered sky background is $ m_{\rm sky} \simeq 20.9$ mag
arcsec$^{-2}$ (corresponding to a Moon eclipse), so that the
typical sky luminosity is $N_{\rm sky} \sim 1600$ counts/pixel,
which enters in the baseline count estimates. However, for
comparison purposes with \cite{Kerins04}, some results in Tables
\ref{table1}, \ref{table2} and \ref{table3} are also given for a
sky background $m_{\rm sky} \simeq 19.5$ mag arcsec$^{-2}$ and
$f_{\rm see} = 0.40$ (corresponding to a randomly positioned PSF
within the superpixel).

Moreover, all the figures presented in Section 6 are given
for the Reference model (see Table \ref{table0}).
The effect of varying the model parameters
for the M31 bulge, disk and halo is also shown in Tables
\ref{table1}-\ref{table3}.

Finally, we recall that $N_{\rm bl}(x,y)$ is obtained from eq.
(\ref{Nbl}) where $N_{\rm gal}(x,y)$ follows from the M31
photometry given by Kent (\citeyear{Kent}).

\subsection{Threshold impact parameter}

In Fig. \ref{fig1} maps of the mean threshold impact parameter
$\langle u_T(x,y) \rangle _{\phi}$  and $\langle u_T^{~2}(x,y)\rangle_{\phi}$
towards M31 are shown.

In this and following figures we use Cartesian coordinates $x$ and
$y$ centered on M31 and aligned along the major and minor axes of
the projected light profile, respectively.

As one can see in Fig. \ref{fig1}, the effect of the higher
luminosity of the inner region of M31 with respect to the outer
part of the galaxy is to reduce the obtained  $\langle
u_T(x,y)\rangle _{\phi}$ values by about an order of magnitude.

In Fig. \ref{fig0}, for selected lines of sight to M31, we show
how $\langle u_T(x,y) \rangle _{\phi}$ depends on the photon
counts $N_{\rm sky}$ from the background, which is approximated as
a diffuse source of magnitude $m_{\rm sky}$ in the range $ 20.9 -
18.9$  mag arcsec$^{-2}$. In Fig. \ref{fig0} we consider several
lines of sight to M31 with different $(x,y)$ coordinates (in units
of arcmin) in the orthogonal plane. Red lines, from the bottom to
the top, refer to $(0,-0.2)$, $(4,-0.2)$ and $(8,-0.2)$, blue
lines are for $(0,-2)$, $(4,-2)$ and $(8,-2)$ coordinates. It is
evident that $\langle u_T(x,y) \rangle _{\phi}$ weakly depends on
$N_{\rm sky}$ in the inner M31 regions, where $N_{\rm bl}$ is
dominated by the counts  $N_{\rm gal}$ from the M31 surface
brightness. Moreover, for a fixed number of counts $N_{\rm sky}$
from the sky, $\langle u_T(x,y) \rangle _{\phi}$ decreases with
increasing $N_{\rm gal}$.

We note that by averaging $\langle u_T(x,y)\rangle _{\phi}$ and
$\langle u_T^{~2} (x,y)\rangle _{\phi}$ (weighting with the star
number density) on the whole field of view towards the M31 galaxy,
we obtain $\langle u_T\rangle_{\phi} \simeq 5.10 \times 10^{-2}$
and $\langle u_T^{~2} \rangle _{\phi} \simeq 8.8 \times 10^{-3}$.

In Table \ref{table00} the effect on $\langle u_T(x,y)\rangle
_{\phi}$ and $\langle u_T^{~2} (x,y)\rangle _{\phi}$ of changing
the parameter values for the superpixel dimension N$\times$N,
$m_{\rm sky}$, $f_{\rm see}$ and $\sigma_T$ is shown.  This is
relevant since, referring to the subsequent Tables
\ref{table1}-\ref{table3}, one can verify that the results for the
Reference model (in the last two rows) scale with $\langle
u_T\rangle_{\phi}$ (in Tables \ref{table1} and \ref{table2}) and
$\langle u_T^{~2}\rangle_{\phi}$ (last four rows in Table
\ref{table3}). Therefore, since $\langle u_T(x,y)\rangle _{\phi}$
and $\langle u_T^{~2} (x,y)\rangle _{\phi}$ strongly depend on the
above mentioned parameters, we expect that all pixel lensing
estimated quantities heavily depend on the observing conditions
and telescope capabilities.

\begin{table*}
\caption{ The threshold impact parameters $\langle u_T \rangle
_{\phi}$ and $\langle u_T^{~2} \rangle _{\phi}$ averaged over the
whole M31 galaxy are given for different values of the superpixel
dimension N$\times$N, sky background $m_{\rm sky}$, fraction
$f_{\rm see}$ of the superpixel covered by the PSF and superpixel
flux stability $\sigma_T$. }
\medskip
\begin{tabular}{|cccc|c|c|}
\hline
N$\times$N &$m_{\rm sky}$&$f_{\rm see}$&
$\sigma_T$&$\langle u_T \rangle _{\phi}$&$\langle u_T^{~2} \rangle _{\phi}$ \\
             & (mag arcsec$^{-2}$)       &          &                  &  &\\
\hline
3 x 3 &20.9&0.87&$2.5\times 10^{-3}$ N$_{\rm bl}$ &$1.10\times 10^{-1}$&$3.17\times 10^{-2}$\\
\hline
5 x 5 &20.9&0.87&$2.5\times 10^{-3}$ N$_{\rm bl}$ &$7.11\times 10^{-2}$&$1.59\times 10^{-2}$\\
\hline
7 x 7 &20.9&0.87&$2.5\times 10^{-3}$ N$_{\rm bl}$ &$5.10\times 10^{-2}$&$9.56\times 10^{-3}$\\
\hline
7 x 7 &20.9&0.40&$2.5\times 10^{-3}$ N$_{\rm bl}$ &$2.57\times 10^{-2}$&$3.13\times 10^{-3}$\\
\hline
7 x 7 &19.5&0.40&$2.5\times 10^{-3}$ N$_{\rm bl}$ &$1.44\times 10^{-2}$&$1.17\times 10^{-3}$\\
\hline
7 x 7 &19.5&0.40&$1.0\times 10^{-2}$ N$_{\rm bl}$ &$3.82\times 10^{-3}$&$1.23\times 10^{-4}$\\
\hline
\end{tabular}
\label{table00}
\end{table*}

\subsection{Pixel lensing optical depth}

Classical microlensing optical depth maps for selected M31 source and lens
populations are shown in Fig. \ref{fig2} for the Reference model.

Here and below, sources and lenses in the M31 bulge and disk are
indicated by indices 1 and 2, while lenses in the M31 halo and MW
disk and halo by indices 3, 5 and 6. The first and second indices
refer to source and lens, respectively.

As one can see, $\langle \tau_c(x,y)\rangle $ always increases
towards the M31 center. The well-known far-to-near side asymmetry
of the M31 disk is clearly demonstrated in $\langle
\tau_c(x,y)\rangle _{23}$, where the lenses are in the M31 halo.
Moreover, a strong asymmetry in the opposite direction in the
bulge-disk (12) and disk-bulge (21) events (due to the relative
source-lens location) is also evident.

We have also found that the classical mean optical depth $\langle
\tau_c(x,y)\rangle $
for lenses in our Galaxy ($\langle \tau_c(x,y)\rangle _{15}$,
$\langle \tau_c(x,y)\rangle _{25}$,
$\langle \tau_c(x,y)\rangle _{16}$ and $\langle \tau_c(x,y)\rangle _{26}$)
is almost constant in any direction and therefore we do not show
the corresponding maps. For reference, the obtained values are
$\langle \tau_c\rangle _{15} \simeq \langle \tau_c\rangle _{25}
\simeq 0.03 \times 10^{-6}$ and
$\langle \tau_c\rangle  _{16} \simeq \langle \tau_c\rangle _{26}
\simeq 0.23 \times 10^{-6}$.

\begin{table*}
\caption{The expected number of events $\langle E_p \rangle$ per
year in pixel lensing observations towards the M31 galaxy for
different locations of sources and lenses is shown. We consider
the 100x70 arcmin$^2$ region oriented along the major axis of M31.
and exclude events occurring within a radius of 8 arcmin of the
M31 center. Sources and lenses in the M31 bulge and disk are
indicated by indices 1 and 2, while lenses in the M31 halo and MW
disk and halo by indices 3, 5 and 6. The first and second indices
refer to source and lens, respectively. The mean mass of bulge and
disk stars is $\sim 0.31~M_{\odot}$ and $\sim 0.53~M_{\odot}$,
respectively. For the lenses in the M31 and MW halos we take a
mass of $\simeq 0.5~M_{\odot}$ and a MACHO fraction $f_{MACHO}
\simeq 20\%$. }
\medskip
\begin{tabular}{c|c|cccccccccc|c}
\hline
             &     & 11 & 12 & 13 & 15 & 16 & 21 & 22 & 23 & 25 & 26 & overall \\
\hline
Model        &$\langle u_T \rangle_\phi$        &    &    &    &    &    &    &    &    &    &   & \\
\hline
Massive halo &$5.10\times 10^{-2}$&11.9&16.0&   10.7&    0.8&    4.2&   15.9&   11.7&   25.4& 1.7&  9.3& 107.5\\
Massive bulge&   &   57.9 & 100.9&10.4&1.6&    9.0&   20.8&   20.5&    8.4& 1.1&  5.9& 236.4\\
Massive disk &   &   11.8 &  92.8&3.7 &0.7&    4.2&   16.1&   69.3&   10.2& 1.7&  9.4& 219.9\\
Reference    &   &   12.5 & 26.4 &8.9 &0.8&    4.4&   13.4&   16.0&   19.5& 1.5&  8.3& 111.8\\
\hline
\hline
Reference    &$1.4\times 10^{-2}$&3.5 &7.2&    2.4&    0.2&    1.2&    3.6& 4.3&  5.2& 0.4&  2.2& 30.4 \\
\hline
\end{tabular}
\label{table1}
\end{table*}
In Fig. \ref{fig3}a classical optical depth maps towards M31 are
given for self-lensing ($\tau_{self} =
\tau_{11}+\tau_{12}+\tau_{21}+\tau_{22}$) and dark-lensing
($\tau_{dark} = \tau_{13}+\tau_{16}+\tau_{23}+\tau_{26}$). The
total contribution $\tau_{tot} = \tau_{self}+\tau_{dark}$ is given
at the bottom of the same figure.

We notice that, in order to evaluate $\tau_{self}$ and $\tau_{dark}$,
we sum optical depths obtained for different source
populations and therefore the averaging procedure
in eq. (\ref{tauposition}) is done
by normalizing with the factor $\int [n_1(D_s;x,y)+n_2(D_s;x,y)] dD_s$.
Mean pixel lensing optical depth $\langle \tau_p(x,y)\rangle $
maps are shown in Fig. \ref{fig3}b.

As we can see, the main effect of the threshold impact parameter
is to substantially decrease $\langle \tau_p(x,y)\rangle$ (with
respect to $\langle \tau_c(x,y)\rangle $ values) in particular
towards the central regions of M31, as a consequence of the
increasing luminosity. Indeed, on average $\langle \tau_p \rangle
\simeq \langle u_T^{~2} \rangle _{\phi} \langle \tau_c \rangle$
and $\langle u_T^{~2} \rangle _{\phi} \simeq 10^{-2}$ (see the
third row in Table \ref{table00}) for the parameter values used in
the Figures.

\subsection{Pixel lensing rate and expected event number}

Maps of the expected number of events in pixel lensing surveys
towards M31 are shown in Fig. \ref{fig5} for the Reference model.
As for the optical depth, we give the number of events separately
for self-lensing, dark-lensing and also the total contribution.

\begin{table*}
\caption{ \bf The same as in Table \ref{table1} for lenses located
in the M31 galaxy. In the last three columns we give the
calculated pixel event number for the South/North Semisphere and
in brackets their ratio. }
\medskip
\begin{tabular}{c|c|ccc|c|ccc}
\hline
Model &$\langle u_T \rangle_{\phi}$&bulge &disk &M31halo & M31overall& stellar      &    M31halo     & 23 \\
\hline
Massive halo &$5.10\times 10^{-2}$& 27.7&27.7&36.1&91.5       &31.4/24.0(1.3)& 25.3/10.8(2.3)&  19.6/5.8(3.4) \\
Massive bulge&                   & 78.6&121.4&18.7&218.7     &72.6/127.4(0.6)&11.3/7.4(1.5)&6.0/2.4(2.5) \\
Massive disk &                   & 27.9&162.0&13.9&203.8      &85.5/104.5(0.8)&9.2/4.7(1.9)&7.2/2.9(2.5)\\
Reference    &                   & 25.9&42.4&28.4&96.7        &33.9/34.4(1.0)&19.6/8.9(2.2)&14.9/4.6(3.2)\\
\hline
\hline
Reference    &$1.4\times 10^{-2}$&  7.1&11.5& 7.7&26.3        &9.3/9.4(1.0)  &5.3/2.4(2.2)& 4.0/1.2(3.2) \\
\hline
\end{tabular}
\label{table2}
\end{table*}

In Figs. \ref{fig5}a and \ref{fig5}b we show, as a function of
position, maps of the instantaneous event number density $\langle
IE_p(x,y) \rangle$ (events per arcmin$^{2}$) and the event rate
$\langle E_p(x,y) \rangle$  (events per year and arcmin$^{2}$).

or the optical depth, the effect of the threshold impact parameter
is to produce a decrease of the event number density towards the
M31 center (for $r \ut <  2 $ arcmin) and an overall reduction of
the event number density with respect to the expectations from
classical microlensing results. Moreover, in the figures it is
also evident that the inner region (within about 10 arcmin from
the M31 center) is dominated by self-lensing events.

In Fig. \ref{f8f8}  the projected (along the $x$ axis) mean event
number density $\langle E_p(y) \rangle $ as a function of the
coordinate $y$ is given. The dashed line refers to dark lensing
events by MACHOs in M31 and MW halos while the solid line is for
self-lensing events by stars in M31 bulge and disk. The
North/South asymmetry is evident for dark events that are
relatively more numerous in the South Semisphere, corresponding to
the far side of the M31 disk.

In Table \ref{table1}, for selected locations of sources
(stars in M31 bulge and disk) and lenses
(stars in M31 bulge and disk, stars in MW disk
and MACHOs in M31 and MW halos)
we give the expected total number of events detectable
by monitoring for 1 year the 100x70 arcmin$^2$ region oriented along the
major axis of M31 (events within 8 arcmin from the center are excluded).
The first four lines refer to the models considered in Table \ref{table0}
and to the parameters in the third row of Table \ref{table00}.
As one can see, the obtained results for the Reference model
are intermediate with respect to those for the other more extreme models.

In the last row of Table \ref{table1}, for the Reference model we
show how the expected event number changes considering a different
value of $\langle u_T \rangle_{\phi} \simeq 1.44 \times 10^{-2}$
(see $5^{\rm th}$ row in Table \ref{table00}). As expected, one
can verify that roughly the event number scales as $\langle u_T
\rangle_{\phi}$.

Similar results have been obtained in previous simulations (see,
e.g. \cite{Kerins04} and references therein). We also note that
our numerical results scale with the fraction of halo dark matter
in form of MACHOs and with the MACHO mass by a factor
$(f_{MACHO}/0.2)~\sqrt{0.5~M_{\odot}/m_l}$.

In Table \ref{table2} we give the total event number $\langle E_p
\rangle$ for different lens populations (bulge, disk and halo)
located in M31. As one can see, the ratio dark/total events
depends on the considered model, varying from 0.07 (for the
massive disk model) to 0.40 for the massive halo model.

To study the far-disk/near-disk asymmetry, in the last three
columns of Table \ref{table2} we give results for the South/North
M31 Semispheres and in brackets their ratio. For the Reference
model, we find that self-lensing events are roughly symmetric (the
same is true for lenses located in the MW disk and halo, not given
in the table), while events due to lenses in M31 halo are
asymmetrically distributed with a ratio of about 2. The asymmetry
is particularly evident (in the last column of the table) for
sources located in the disk.

In Table \ref{table3} the instantaneous total number of events
$\langle IE_p \rangle$ within the considered M31 region is given.
The first four rows refer to the parameter values $\langle m_b
\rangle \simeq 0.31 ~M_{\odot}$, $\langle m_d \rangle \simeq 0.53
~M_{\odot}$, $f_{MACHO} = 0.2$ and $\langle u_T^{~2}
\rangle_{\phi} \simeq 9.56 \times 10^{-3}$ (used throughout the
paper). For comparison with the results obtained by
\cite{Kerins04}, in the last four rows of Table \ref{table3} we
present our results for $\langle m_b \rangle \simeq 0.5
~M_{\odot}$, $\langle m_d \rangle \simeq 0.5 ~M_{\odot}$,
$f_{MACHO} = 1$ and $\langle u_T^{~2} \rangle_{\phi} \simeq 1.17
\times 10^{-3}$. The asymmetry ratio we obtain is always rather
smaller than that quoted by \cite{Kerins04}.

As it has been mentioned by several authors, in order to
discriminate between self and dark lensing events, it is important
to analyze the event duration. Indeed self-lensing events are
expected to have, on average, shorter duration with respect to
events due to halo MACHOs.

\subsection{Pixel lensing event time scale}

Maps of mean event duration time scale in classical and pixel lensing are
shown in Fig. \ref{fig6}a and \ref{fig6}b.

Here we use the probability, for each location of sources and
lenses given in eq. (\ref{pp}), of obtaining event duration maps
for self and dark microlensing events.

As expected, short duration events are mainly distributed towards
the inner regions of the galaxy and this occurs for both  $\langle
t_c(x,y)\rangle$ and $\langle t_p(x,y)\rangle $. The main effect
of $\langle u_T(x,y)\rangle _{\phi}$ is to decrease the event time
scale, in particular towards the inner regions of M31, giving a
larger number of short duration events with respect to
expectations based on $\langle t_c(x,y)\rangle $ calculations.

Both for self and dark events the pixel lensing time scale we
obtain is $\simeq 1 - 7$ days, in agreement with results in
\cite{Kerins04}, but much shorter with respect to the duration of
the events observed by the MEGA Collaboration \citep{Mega04}. This
is most likely due to the fact that current experiments may not
detect events shorter than a few days.

However, the pixel lensing time scale values depend on $\langle
u_T(x,y) \rangle _{\phi}$ and ultimately on the observational
conditions and the adopted analysis procedure. Indeed from Table
\ref{table00} one can see that the $\langle u_T(x,y) \rangle
_{\phi}$ value may be easily doubled, changing the adopted
parameters and therefore giving longer events.

In Fig. \ref{f9f9} the pixel lensing event duration
$\langle t_p(y) \rangle $
averaged along the $x$ direction
is given as a function of the $y$ coordinate.
The dashed line refers to dark lensing events by MACHOs in M31 and MW halos
while the solid line is for self-lensing events by stars in M31 bulge and disk.

It is clearly evident that dark events last roughly twice as long
as self-lensing events and that the shortest events are expected
to occur towards the M31 South Semisphere.

The presence of a large number of short duration events in pixel
lensing experiments towards M31 has been reported by several
authors \citep{Paulin03,Paulin04}.

\begin{table*}
\caption{The instantaneous number of events in pixel lensing
observations towards the M31 galaxy for different locations of
sources and lenses is shown (for details see text). Numbers in
brackets refer to the South Semisphere of M31. For the MW disk and
halo, lenses in the South Semisphere of the MW contribute to
roughly one half of the total and so the corresponding event
numbers are not given. }
\medskip
\begin{tabular}{c|c|ccc|c|ccc}
\hline
Model &$\langle u_T^{~2}\rangle_{\phi}$&bulge &disk &M31halo & M31overall& stellar      &    M31halo     & 23 \\
\hline
Massive halo &$9.56\times 10^{-3}$& 0.74 & 0.57&  2.25&    3.56&  0.68/0.63(1.08)&  1.52/0.73(2.09)&  1.16/0.42(2.74) \\
Massive bulge&                    & 2.22 & 2.87&  1.37&    6.45&  1.74/3.34(0.52)&  0.83/0.54(1.55)&  0.41/0.17(2.46) \\
Massive disk &                    & 0.74 & 3.35&  0.98&    5.07&  1.32/2.77(0.48)&  0.64/0.34(1.86)&  0.49/0.21(2.29) \\
Reference    &                    & 0.70 & 0.90&  1.78&    3.37&  0.69/0.90(0.77)&  1.21/0.57(2.11)&  0.89/0.31(2.89) \\
\hline
\hline
Massive halo &$1.17\times 10^{-3}$& 0.08& 0.05&  1.34&    1.47&  0.09/0.05(1.84)&  0.95/0.39(2.47)&  0.85/0.30(2.82) \\
Massive bulge&                    & 0.20& 0.20&  0.67&    1.08&  0.18/0.22(0.82)&  0.44/0.23(1.88)&  0.32/0.13(2.50) \\
Massive disk &                    & 0.09& 0.29&  0.58&    0.96&  0.17/0.21(0.81)&  0.39/0.19(2.10)&  0.35/0.15(2.32) \\
Reference    &                    & 0.08& 0.07&  1.04&    1.19&  0.09/0.07(1.32)&  0.74/0.29(2.52)&  0.66/0.22(2.96) \\
\hline
\end{tabular}
\label{table3}
\end{table*}

\section{Conclusions}

We have studied the optical depth, event number and time scale
distributions in pixel lensing surveys towards M31 by addressing,
in particular, the changes with respect to expectations from
classical microlensing (in which the sources are resolved).

Assuming, as reference values, the capabilities of the Isaac
Newton Telescope in La Palma and typical CCD camera parameters,
exposure time and background photon counts, we perform an analysis
consisting of averaging all relevant microlensing quantities over
the threshold value $u_T(x,y)$ of the impact parameter. Clearly,
as in classical microlensing estimates, an average procedure is
also done with respect to all the other parameters entering in
microlensing observables: source and lens position, lens mass and
source and lens transverse velocities.

The M31 bulge, disk and halo mass distributions are described
following the Reference model in \cite{Kerins04}, which provides
remarkably good fits to the M31 surface brightness and rotation
curve profiles. We also take a standard mass distribution model
for the MW galaxy, as described in Section 2, and assume that M31
and MW halos contain 20\% 0.5 $M_{\odot}$ MACHOs.

We consider red giants as the sources that most likely may be
magnified (and detected in the red band) in microlensing surveys.
Moreover, given the lack of precise information about the stellar
luminosity function in M31, we assume that the same function holds
both for the Galaxy and M31 and does not depend on the position.
Accordingly, the fraction of red giants (over the total star
number) is $\simeq 5.3 \times 10^{-3}$.

Our main results are maps in the sky plane towards M31
of threshold impact parameter
$\langle u_T(x,y) \rangle _{\phi}$,
optical depth $\langle \tau_p(x,y)\rangle $,
instantaneous event number density $\langle IE_p(x,y)\rangle$
(events per arcmin$^2$ of ongoing microlensing events at any instant
in time)
and event number density $\langle E_p(x,y)\rangle $
(events per yr and arcmin$^2$ to be detected in M31 surveys)
and time scale $\langle t_p(x,y)\rangle$.

These maps show an overall reduction of the corresponding classical
microlensing results and also a distortion of their shapes with respect
to other results in the literature.

Figs. \ref{fig2} and \ref{fig3} show maps of the mean optical
depth (averaged over the source number density) for the different
source and lens locations.

In Fig. \ref{fig5} we give the instantaneous pixel lensing
event number density and the event rate for self, dark and total
lensing. It clearly appears that the central region of M31
is dominated by self-lensing events due to sources and lenses in M31 itself,
while dark events are relatively more numerous in the outer region
(see also Fig. \ref{f8f8}).

In Tables \ref{table1}-\ref{table2}, for the M31 mass distribution
models considered by \cite{Kerins04}, we give the expected total
event number $\langle E_p \rangle$ to be detected by monitoring,
for 1 yr, the 100x70 arcmin$^2$ region oriented along the major
axis of M31 (the inner 8 arcmin region is excluded). We find that
the expected dark to total event number ratio is between 7\% (for
the massive disk model) and 40\% (for the massive halo model). The
tables also show the well-known far-disk/near-disk asymmetry due
to lenses in the M31 halo. Self-lensing events, instead, are
distributed more symmetrically between the M31 North and South
Semisphere. Similar conclusions are evident from Table
\ref{table3}, where we give the instantaneous number of events,
although the asymmetry ratio we obtain is always smaller than the
values quoted by \cite{Kerins04}.

Fig. \ref{fig6}) shows a decrease of the event time scale with
respect to classical microlensing, particularly towards the inner
regions of M31, due to the high brightness of the galaxy. Both for
self and dark lensing events, the pixel lensing time scale we
obtain is $\simeq 1 - 7$ days, in agreement with results in the
literature. Note that the duration of the events observed by the
MEGA Collaboration \citep{Mega04} is typically much longer than 7
days, due to the difficulty of detecting short events in current
experiments. It is also clear from Fig. \ref{fig6} that dark
events last roughly twice as long as self-lensing events and that
the shortest events are expected to occur towards the M31 South
Semisphere (see Fig. \ref{f9f9}).

However, we emphasize that the pixel lensing results obtained
depend on $\langle u_T(x,y) \rangle _{\phi}$ and $\langle
u_T^{~2}(x,y) \rangle _{\phi}$ values, and ultimately on the
observing conditions and telescope capabilities. Indeed, from
Table \ref{table00}, where the values of $\langle u_T \rangle
_{\phi}$ and $\langle u_T^{~2} \rangle _{\phi}$ averaged over the
whole M31 galaxy are given, one can verify that pixel lensing
quantities scaling with $\langle u_T \rangle _{\phi}$ ($\langle
E_p \rangle$ and $\langle t_p \rangle$) may vary by more than one
order of magnitude while quantities scaling with $\langle u_T^{~2}
\rangle _{\phi}$ ($\langle \tau_p \rangle$ and $\langle IE_p
\rangle$) may change by two orders of magnitude.

The present analysis can be used to test estimates and Monte-Carlo
simulations by other Collaborations and it has also been performed
in view of a planned survey towards M31 by the SLOTT-AGAPE
Collaboration \citep{Slott00}.

\begin{acknowledgements}
We acknowledge S. Calchi Novati, Ph. Jetzer and F. Strafella
for useful discussions.

\end{acknowledgements}

\end{document}